\documentclass[aps,prl,twocolumn,groupedaddress,showpacs,preprintnumbers,superscriptaddress,
amsmath,amssymb]{revtex4}
\usepackage[dvips]{graphicx}
\usepackage{amsfonts}
\usepackage{amssymb}
\usepackage{float}
\usepackage{natbib}
\usepackage{amsmath}
\usepackage{float}
\usepackage{dcolumn}
\usepackage{lscape}
\usepackage{psfrag}         

\newcommand{\beq}{\begin{equation}}
\newcommand{\eeq}{\end{equation}}
\newcommand{\bfg}{\begin{figure}}
\newcommand{\efg}{\end{figure}}

\newcommand{\bea}{\begin{eqnarray}}
\newcommand{\eea}{\end{eqnarray}}

\makeatletter
\newlength{\earraycolsep}
\def\eqnarray{\stepcounter{equation}\let\@currentlabel%
\theequation
\global\@eqnswtrue\m@th
\global\@eqcnt\z@\tabskip\@centering\let\\\@eqncr
$$\halign to\displaywidth\bgroup\@eqnsel\hskip\@centering
$\displaystyle\tabskip\z@{##}$&\global\@eqcnt\@ne
\hskip 2\earraycolsep \hfil$\displaystyle{##}$\hfil
&\global\@eqcnt\tw@ \hskip 2\earraycolsep
$\displaystyle\tabskip\z@{##}$\hfil
\tabskip\@centering&\llap{##}\tabskip\z@\cr}
\makeatother

\begin{document}

 \title{Semiclassical Mechanism for the Quantum Decay in Open Chaotic Systems}

 \author{Daniel Waltner}
\affiliation{Institut f\"ur Theoretische Physik, Universit\"at Regensburg, D-93040 Regensburg, Germany}
 \author{ Martha Guti\'errez}
\affiliation{Institut f\"ur Theoretische Physik, Universit\"at Regensburg, D-93040 Regensburg, Germany}
 \author{Arseni Goussev}
\affiliation{Institut f\"ur Theoretische Physik, Universit\"at Regensburg, D-93040 Regensburg, Germany}
\affiliation{School of Mathematics, University of Bristol, University Walk, Bristol BS8 1TW,
United Kingdom}
 \author{Klaus Richter}
\affiliation{Institut f\"ur Theoretische Physik, Universit\"at Regensburg, D-93040 Regensburg, Germany}

\date{\today}

\begin{abstract} 
We address the decay in open chaotic quantum systems and calculate
semiclassical corrections to the classical exponential decay. We
confirm random matrix predictions and, going beyond,
calculate Ehrenfest time effects. To support our results we
perform extensive numerical simulations. Within our approach we
show that certain (previously unnoticed) pairs of interfering,
correlated classical trajectories are of vital
importance.
They also provide the dynamical mechanism
for related phenomena such as photo-ionization and -dissociation,
for which we compute cross section correlations.
Moreover, these orbits allow us to establish a semiclassical
version of the continuity equation.
\end{abstract}

\pacs{03.65.Sq,05.45.Mt, 05.45.Pq }
\maketitle


Besides their relevance to many areas of physics, open quantum systems
play an outstanding role in gaining an improved understanding of the
relation between classical and quantum physics \cite{JPA}. 
For a closed quantum system
the spatially integrated probability density
 \beq\label{eq1}
 \rho\left(t\right) = \int_V d{\bf r}\, \psi({\bf r},t)\psi^*({\bf r},t) 
 \eeq
 of a wave function $ \psi({\bf r},t) $ in the volume $V$
is constant, {\rm i.e.} $\rho (t)\!\equiv\! 1$. This fact is 
 naturally retained when taking the classical 
limit  in a semiclassical evaluation of Eq.\,(\ref{eq1}),
reflecting particle conservation in the quantum and classical limit.  However,
when opening up the system, $\rho(t)$, then representing the quantum 
survival probability, exhibits deviations from its classical counterpart $\rho_{\rm cl}(t)$;
in other words, certain quantum properties of the closed system can be unveiled upon 
opening it.

For an open quantum system with a classically chaotic coun\-terpart,
the classical survival probability is asymptotically $\rho_{\rm
  cl}(t)\!=\! \exp{(-t/ \tau_d) } $, with classical dwell time $\tau_d$.
This has been observed in various disciplines, either directly, as in
atom billiards \cite{ref:Raizen01,ref:Friedman01}, or indirectly in
the spectral regime of Ericson fluctuations in electron
\cite{ref:Marcus92} or microwave \cite{ref:Doron90} cavities, and in
atomic photo-ionization \cite{ref:Stania05}.

However, it was found numerically \cite{ref:Casati} and confirmed with
supersymmetry techniques \cite{ref:Frahm97,ref:Savin} that the difference between $\rho(t)$ and $\rho_{\rm cl}(t)$ becomes significant at times close to the quantum relaxation time $t^* = \sqrt{\tau_d t_H}$. In the semiclassical limit $t^*$ is shorter than the
Heisenberg time $t_H = 2\pi\hbar / \Delta$ (with $\Delta$ the mean
level spacing). It was shown in
Refs.~\cite{ref:Frahm97,ref:Savin} that $\rho(t)$ is a universal function
depending only on $\tau_d $ and $t_H$ in the random matrix
theory (RMT) limit \cite{ref:Kaplan}.  Though the leading quantum deviations from $\rho_{\rm cl}(t)$ were reproduced semiclassically for quantum
graphs \cite{ref:Puhlmann05}, a general understanding of its dynamical
origin is still lacking.

In this Letter we present a semiclassical calculation of $\rho(t)$ for $t<t^*$ for
general, classically chaotic systems. It reveals the mechanism
underlying the appearance of quantum
corrections upon opening the system.
In our calculation we go beyond the so-called diagonal approximation
and evaluate contributions from correlated trajectory pairs
\cite{ref:loops}.  This technique has been extended and applied to
calculate various spectral \cite{ref:Heusler,ref:Brouwer06B} and
scattering
\cite{ref:Richter02,ref:Adagideli03,ref:Ehrenfest2,ref:Heusler06,ref:Brouwer06,ref:Jacquod06,ref:Kuipers08}
properties of quantum chaotic systems.
We find however that for calculating $\rho(t)$ a new class of
correlated trajectory pairs, `one-leg-loops', has to be considered
along with the previously known loop diagrams. They prove particularly
crucial for ensuring unitarity in problems involving semiclassical
propagation along open trajectories inside a system and, moreover,
allow one to semiclassically recover the continuity equation.

We present the dominant quantum corrections to $\rho_{\rm cl}(t)$ for systems with and without time reversal
symmetry. Going beyond RMT, we calculate Ehrenfest time effects on
$\rho(t)$, compare with quantum simulations of billiard
dynamics, and extend our approach to photo-ionization and
-dissociation cross sections.

{\it Semiclassical approach --}
We consider $\rho(t)$, Eq.\,(\ref{eq1}), for a two-dimensional system of area $A$
and  express
$ \psi({\bf r},t)= \int d{\bf r'} K({\bf r},{\bf r'};t)\psi_0({\bf r'})  $
through the initial wave function $\psi_0({\bf r}')$ 
and the time-dependent propagator 
$ K({\bf r},{\bf r'};t)$  that we
approximate  semiclassically by \cite{ref:Gutzwiller90}
\beq\label{eq:K}
K_{\rm sc}\left(\mathbf{r},\mathbf{r'};t\right)=
\frac{1}{2\pi i\hbar} \sum_{\bar{\gamma}\left(\mathbf{r'}\to\mathbf{r},t\right)}
\! D_{\bar{\gamma}} {\rm e}^{i S_{\bar{\gamma}}/\hbar} \, .
\eeq
Here $S_{\bar{\gamma}}\!=\!S_{\bar{\gamma}}({\bf r},{\bf r'};t) $
is the classical action along the path $\bar{\gamma}$ connecting ${\bf r}'$ 
and ${\bf r}$ in time $t$, and
$D_{\bar{\gamma}}\!=\!\left|{\rm det}(\partial^2S_{\bar{\gamma}} / \partial {\bf r} \partial {\bf r'})
\right|^{1/2}\!\! \!\!\exp(-i\pi\mu_{\bar{\gamma}}/2)$ with  Maslov index $\mu_{\bar{\gamma}}$.

The semiclassical survival probability, $\rho_{\rm sc}(t)$, obtained by 
expressing the time evolution of $\psi({\bf r},t)$ and $\psi^\ast({\bf r},t)$
in Eq.\,(\ref{eq1}) through
$K_{\rm sc}$, Eq.\,(\ref{eq:K}), 
is given by three spatial integrals over a double sum over trajectories
$\bar{\gamma}, \bar{\gamma}'$ starting at initial 
points $\mathbf{r}'$ and $\mathbf{r}''$, weighted by $\psi_0(\mathbf{r}')$ and
$\psi_0^\ast(\mathbf{r}'')$, and ending at the same point $\mathbf{r}$ inside $A$. 
For simplicity of presentation we here
assume $\psi_0$ to be spatially localized (e.g.\ a Gaussian wave packet); while
generalizations are given below. Introducing 
$\mathbf{r}_0\!=\!(\mathbf{r}'\!+\!\mathbf{r}'')/2$ and
$\mathbf{q}\!=\!(\mathbf{r}'\!-\!\mathbf{r}'')$, we  replace the 
original paths $\bar{\gamma}, \bar{\gamma}'$, 
by nearby trajectories $\gamma$ and $\gamma'$ connecting $\mathbf{r}_0$ and $\mathbf{r}$ in  time $t$.
Then, upon expanding the action
$S_{\bar{\gamma}}({\bf r},{\bf r'};t) \simeq S_{\gamma}({\bf r},{\bf r}_0;t) 
- \mathbf{q} \mathbf{p}_0^\gamma/2$ (with $\mathbf{p}_0^\gamma$ the initial momentum of 
path $\gamma$) and $S_{\bar{\gamma}'}$ analogously, we obtain
\bea
\nonumber
\rho_{\rm sc}(t)&=& \frac{1}{(2\pi \hbar)^2}\int d{\mathbf r} d{\mathbf r}_0 d{\mathbf
q}\, \psi_0\!\left({\mathbf r}_0+\frac{{\mathbf q}}{2}\right)\psi_0^*\left({\mathbf
r}_0 -\frac{{\mathbf q}}{2}\right)
\\ 
 & & \times  \!\!\!  \sum_{\gamma,\gamma'(\mathbf{r}_0\to\mathbf{r},t)}\!\!\!\!
 D_{\gamma}D_{\gamma' }^* 
{\rm e}^{(i/\hbar) [S_\gamma -S_{\gamma'} -
(\mathbf{p}_0^\gamma+\mathbf{p}_0^{\gamma'})\mathbf{q}/2]} .
\label{eq:rho-sc}
\eea
The  double sums in Eq.\,(\ref{eq:rho-sc}) contain rapidly oscillating 
phases $(S_\gamma \!-\!S_{\gamma'})/\hbar$ which are assumed to vanish
unless $\gamma$ and $\gamma'$ are correlated.
The main, diagonal, contribution to $\rho_{\rm sc}$ arises from pairs $\gamma\!=\!\gamma'$ that,
upon employing a sum rule \cite{ref:Richter02},
yield the classical decay
$\rho_{\rm cl}(t) \!=\!\langle {\rm e}^{-t/ \tau_d}\rangle$
\cite{ergodic}.
Here 
$\langle  F \rangle \!=\! (2\pi \hbar)^{-2}
\int d{\mathbf r}_0 d{\mathbf p}_0 F({\mathbf r}_0,{\mathbf p}_0)\rho_{W}({\mathbf r}_0,{\mathbf p}_0)$,
where $\rho_W({\mathbf r}_0,{\mathbf p}_0)\!=\!\int d{\mathbf q}\psi_0({\mathbf r}_0\!+\! 
{\mathbf q}/2)\psi_0^\ast({\mathbf r}_0\!-\!{\mathbf q/2})  e^{-(i/\hbar){\mathbf q}\cdot {\mathbf p}_0}$
is the Wigner transform of the initial state and 
$\tau_d\!=\!\Omega(E)/2 w p$, with 
$\Omega(E)\!=\!\int d{\mathbf r}d {\mathbf p}\delta(E\!-\!H({\mathbf r},{\mathbf p}))$ 
and $w$ the size of the opening. For chaotic billiards $\tau_d(p)\!=\!m\pi A/w p$. 
For initial states with small energy dispersion 
$\rho_{\rm cl}(t)\! =\!{\rm e}^{-t/\! \tau_d(p_0)}$. 

For systems with time reversal symmetry, leading-order quantum corrections
to $\rho_{\rm cl}(t)$ arise from off-diagonal contributions to the double sum in 
Eq.\,(\ref{eq:rho-sc}), given by pairs of correlated orbits depicted as full and dashed 
line in Fig.\,\ref{fig:loops}(a), as in related semiclassical treatments
\cite{ref:loops,ref:Richter02,ref:Adagideli03,ref:Ehrenfest2,ref:Heusler06,ref:Brouwer06,ref:Jacquod06}.
The two orbits are exponentially close to each
other along the two open `legs' and along the loop \cite{ref:Richter02},
but deviate in the intermediate 
encounter region (box in  Fig.\,\ref{fig:loops}(a)). Its length is
$t_{\rm enc}\!=\!\lambda^{-1}\!\ln (c^2/|su|)$ \cite{ref:Heusler},
where $\lambda$ is the Lyapunov exponent, $c$ is a classical
 constant, and $s$ and $u$ are the stable and unstable coordinates in a 
Poincar\'e surface of section (PSS)  in the encounter region. 
Such `two-leg-loops' (2ll) are based on orbit pairs with 
$S_\gamma \!-\! S_{\gamma'}\!=\! su$ and a density 
$w_{\rm 2ll}(s,u,t)\!=\![t-2t_{\rm enc}(s,u)]^2 / [2\Omega(E)t_{\rm enc}(s,u)]$\cite{ref:Heusler06}.
Invoking the sum rule, the double sum in Eq.\,(\ref{eq:rho-sc}) is replaced by
$\int du\int ds e^{(-t+t_{\rm enc})/ \tau_d}w_{\rm 2ll}(s,u,t) e^{(i/\hbar) su}$.
Here 
 $e^{t_{\rm enc}/ \tau_d}$
accounts for the fact that if the first encounter stretch is inside $A$ 
the second must also be inside $A$. 
This gives the 2ll contribution (Fig.\,\ref{fig:loops}(a)) to $\rho(t)$:
\beq\label{eq:rho-2l}
\rho_{\rm 2ll} (t) =  e^{-t/ \tau_d}\left( 
-2\frac{t}{t_H} + \frac{t^2}{2\tau_d t_H} \right) \; .
\eeq
The linear term in Eq.\,(\ref{eq:rho-2l}) violates unitarity, since it
does not vanish upon closing the system, {\em i.e.\ } as $\tau_d \to
\infty$.
This is cured by considering a {\it new type of diagrams}. These orbit
pairs, to which we refer as `one-leg-loops' (1ll), are characterized
by an initial or final point inside the encounter region
(Fig.\,\ref{fig:loops}(b,c)). They are relevant for open orbits
starting or ending {\em inside} $A$ and hence have not arose in
conductance treatments based on lead-connecting paths, since at an
opening the exit of one encounter stretch implies the exit of the
other one \cite{ref:cite1}.
  
\bfg
\includegraphics[width=0.9\columnwidth,clip=true]{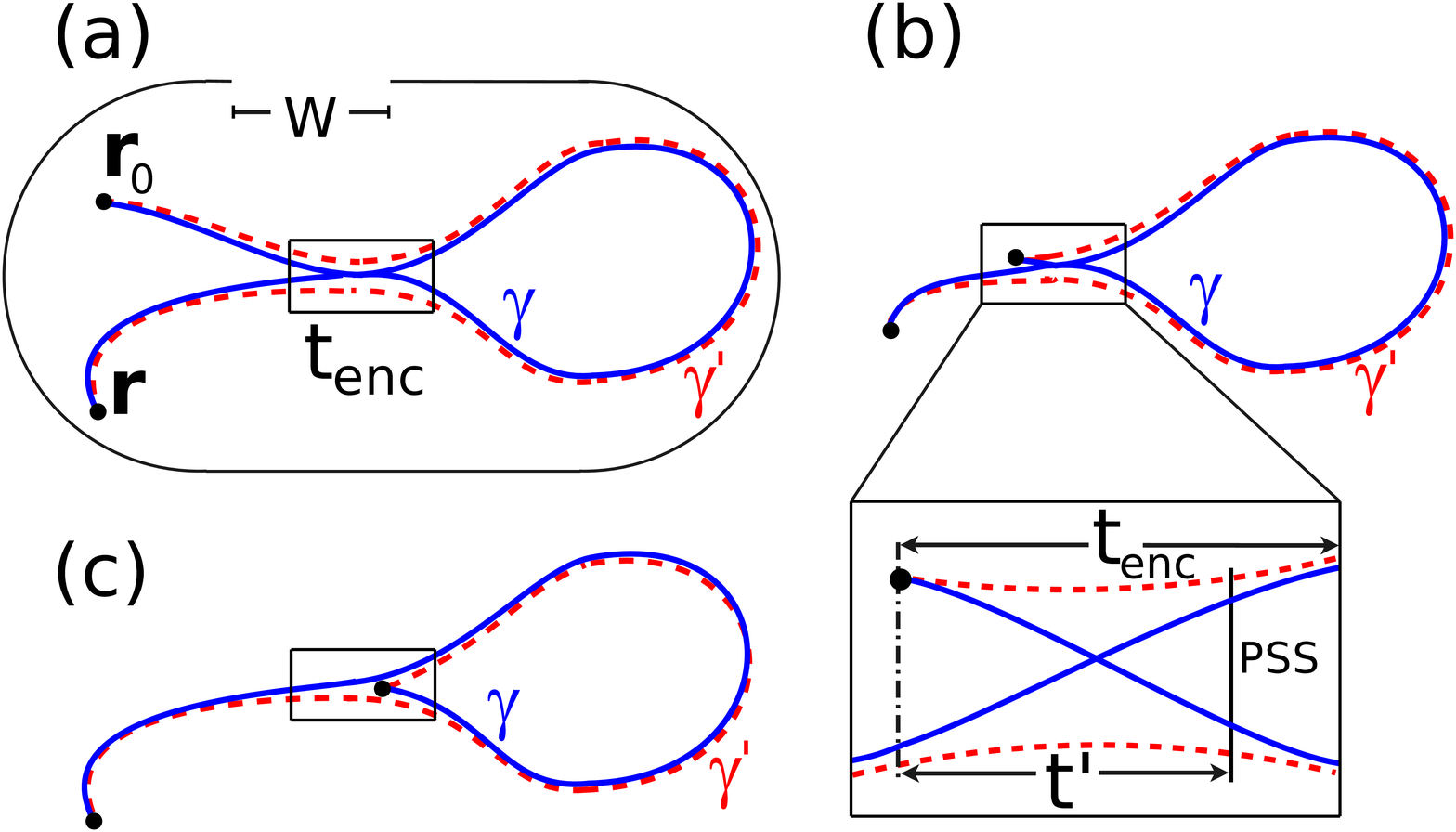}
\caption{(Color online) Pairs of correlated classical trajectories  $\gamma$ (full line)
 and $\gamma'$ (dashed) generating the leading quantum corrections to the classical decay probability. 
While in panel (a) the encounter region (box) connects a loop with two legs, the paths begin or
 end inside the encounter region (`one-leg-loops') in (b) with  and in (c) without 
 a self-crossing in configuration space. The zoom into the encounter region in (b) depicts
 the position of the Poincar\'e surface of section used.}
\label{fig:loops}
\efg

For their evaluation consider 
the time $t'$ between the initial or final point of the trajectory and the PSS, 
defined in the zoom into Fig.\,\ref{fig:loops}(b). Then 
$t_{\rm enc}(t',u) \!=\! =t' \! +\!\lambda^{-1}\! \ln(c/|u|)$ and
$S_\gamma\!-\!S_{\gamma'}\!=\! su$  for any position of the PSS.  The density of 
encounters is $w_{\rm 1ll}(s,u,t)\!=\! 2 \int_{0}^{\lambda^{-1}\ln(c/|s|)} dt'\!  [t\!-\!2t_{\rm 
enc}(t',u)]/[\Omega(E)t_{\rm enc}(t',u)]$,
where the prefactor 2 accounts for the two cases of beginning or ending in an encounter region.
We evaluate this contribution by modifying $\rho_{\rm cl}(t)$ by 
$e^{t_{\rm enc}/ \tau_d}$ as before and integrating over $s$, $u$ and $t'$. To this end
 we substitute \cite{ref:Brouwer06}
$t''\!=\!t'+\lambda^{-1}\ln(c/|u|)$, $\sigma\!=\!c/u$ and $x\!=\!su/c^2$,
with integration domains $-1\!<\!x\!<\!1$, 
$1\!<\!\sigma\!<\!e^{\lambda t''}$ and $0\!<\!t''\!<\! \lambda^{-1}\!\ln(1/|x|)$.
Note that the limits for $t'$ include the case when the paths do not have 
a self-crossing in configuration
space (Fig.\,\ref{fig:loops}(c)). The integration yields
\beq\label{eq:rho-1l}
 \rho_{\rm 1ll}\left(t\right) = 2\frac{t}{t_H}e^{-t/ \tau_d} \, .
\eeq
It precisely cancels the linear term in $\rho_{\rm 2ll}$, Eq.\,(\ref{eq:rho-2l}),
{\em i.e.} 
$ \rho_{\rm 2ll}(t)\!+\!\rho_{\rm 1ll}(t)  \! = \!  e^{-t/ \tau_d}t^2/(2\tau_d t_H)$,
recovering unitarity!

The next-order quantum corrections are obtained by calculating \cite{ref:Gutierrez} 
1ll and 2ll contributions of diagrams such as discussed in \cite{ref:Heusler}. 
Together with Eqs.\,(\ref{eq:rho-2l},\ref{eq:rho-1l}), this yields
for systems with time reversal symmetry
\beq\label{eq:GOE}
\rho_{\rm sc}(t) \simeq
e^{-t/ \tau_d}\! \left(1\!+\!\frac{t^2}{2\tau_d t_H}\!-\!\frac{t^3}{3\tau_d t_H^2}\!+\!
\frac{5t^4}{24\tau_d^2 t_H^2}\right)\, ,
\eeq 
for $t<t^*$. The term quadratic in $t$ represents the weak localization-type enhancement of the quantum
survival probability. The quadratic and the quartic terms, which agree with RMT \cite{ref:Frahm97}, dominate for the time range considered. The cubic term in
Eq.\,(\ref{eq:GOE}), whose functional form was anticipated in
\cite{ref:Frahm97}, scales differently with $\tau_d$ and $t_H$.

 For systems without time reversal symmetry 
 the calculation of the relevant one- and two-leg-loops gives,
again in accordance with RMT \cite{ref:Frahm97},
\beq\label{eq:GUE}
\rho_{\rm sc} (t) \simeq  e^{-t/ \tau_d}\left(1+\frac{t^4}{24\tau_d^2 t_H^2}\right) \, ,
\eeq
We finally note that our restriction to localized initial states can
be lifted and the results generalized to arbitrary initial states
\cite{ref:Gutierrez}. This is because the trajectory pairs
$\bar{\gamma}, \bar{\gamma}'$ survive the integration in
Eq.\,(\ref{eq:rho-sc}) only if their starting points are close to each
other in phase space, rendering the above analysis valid.


{\it Continuity equation.--}
It is instructive to reformulate the decay problem in terms of paths
crossing the opening. 
To this end we consider
the integral version of the continuity equation,
$ \partial \rho ({\mathbf r},t)/\partial t\!+\! \nabla \!\cdot\! \mathbf{j}(\mathbf{r},t) \!=\! 0$,
namely
\beq\label{eq:continuity}
\frac{\partial }{\partial t}\rho (t) = - \int _{S} \mathbf{j}(\mathbf{r},t)\cdot \hat{n}_x\, dx \, ,
\eeq
where $S$ is the cross section of the opening with a normal vector $\hat{n}_x$.
In Eq.\,(\ref{eq:continuity}), the current density 
$ \mathbf{j}(\mathbf{r},t)\!=\!(1/m){\rm Re}[(\hbar/i)\psi^*(\mathbf{r},t)\mathbf{\nabla} 
\psi (\mathbf{r},t)]$ 
can be semiclassically expressed
through Eq.\,(\ref{eq:K}) in terms of orbit pairs connecting  points inside $A$ with the opening.
In the diagonal approximation we obtain 
 $\int _{S}\mathbf{j}_{\rm diag} \cdot
\hat{n}_xdx=e^{-t/\tau_d}/\tau_d $, consistent with $\rho_{\rm
  cl}(t)$.  Loop contributions are calculated analogously to those of
$\rho_{\rm sc}$ from Eq.\,(\ref{eq:rho-sc}), giving 
\beq\label{eq:j-1ll-2ll}
\int_{S} (\mathbf{j}_{\rm 2ll} \!+\! \mathbf{j}_{\rm 1ll})
\cdot \hat{n}_x dx \!=\! e^{-t/\tau_d}\frac{t^2 - 2t\tau_d}{2\tau_d^2 t_H} \, .
\eeq
Time integration of Eq.\,(\ref{eq:continuity}) leads to 
$\rho_{\rm 2ll}(t) \!+\! \rho_{\rm 1ll}(t)\!=\! e^{-t/\tau_d} t^2/(2\tau_dt_H)$,
consistent with Eq.\,(\ref{eq:GOE}).
The 1ll contributions enter into Eq.\,(\ref{eq:j-1ll-2ll}) with half the weight, 
since 1lls with a short leg (encounter box) at the opening must be excluded. 
These 'missing' paths assure the correct form 
of quantum deviations from $\rho_{\rm cl}(t)$.

Higher 2ll and 1ll  corrections to $ \mathbf{j}$ lead 
to Eqs.\,(\ref{eq:GOE},\ref{eq:GUE}). We conclude that both,
2ll {\em and} 1ll contributions to $ \mathbf{j}$ are essential 
to achieve a unitary flow and thereby to establish a semiclassical version of the 
continuity equation.

{\it Ehrenfest time effects.--}
 The Ehrenfest time $\tau_E$ \cite{ref:Chirikov} separates the
 evolution
of wave packets following essentially the classical dynamics from longer time scales dominated
by wave interference.
While $\tau_E$-effects have been mainly considered for stationary processes
involving time integration 
\cite{ref:Aleiner96,ref:Yevtushenko00,ref:Adagideli03,ref:Ehrenfest2,ref:Brouwer06,ref:Jacquod06},
signatures of $\tau_E$ should appear most directly in the time domain 
\cite{ref:Brouwer06B,ref:Schomerus04},
{\em i.e.}  for $\rho(t)$. 
Here we semiclassically compute the $\tau_E$-dependence of the weak-localization correction 
to $\rho(t)$ in Eq.\,(\ref{eq:GOE}).
To this end we distinguish between $\tau_E^{\rm c}\!=\!\lambda^{-1}\!\ln({\mathcal L}/\lambda_B)$,
where ${\mathcal L}$ is the typical system size and $\lambda_B$ the de Broglie wavelength,
and  $\tau_E^{\rm o}\!=\!\lambda^{-1}\!\ln(w^2/({\mathcal L}\lambda_B))$,
related  to the width $w$  of the opening \cite{ref:Jacquod06}.
As before we consider that the densities $w_{\rm 2ll, 1ll}(s,u,t)$ contain the Heaviside function $\theta(t\!-\!2t_{\rm enc})$ (negligible for $\tau_E^{\rm o,c}\ll\tau_d$)  assuring that the time
required to form a 1ll or 2ll is larger than $2t_{\rm enc}$.
Our calculation gives (for $\tau_E^{\rm o,c}\lambda\gg 1$ with $2\tau_E^{\rm e}\!=\!\tau_E^{\rm
o}+\tau_E^{\rm  c}$)
\beq\label{eq:Ehrenfest}
\rho_{\rm 2ll}(t)\!+\!\rho_{\rm 1ll}(t)
=e^{-(t-\tau_E^{\rm o})/\tau_d}\frac{(t-2\tau_E^{\rm e})^2}{2\tau_dt_H}\theta(t-2\tau_E^{\rm e})\, .
\eeq

\bfg
\includegraphics[width=1.0\columnwidth,clip=true]{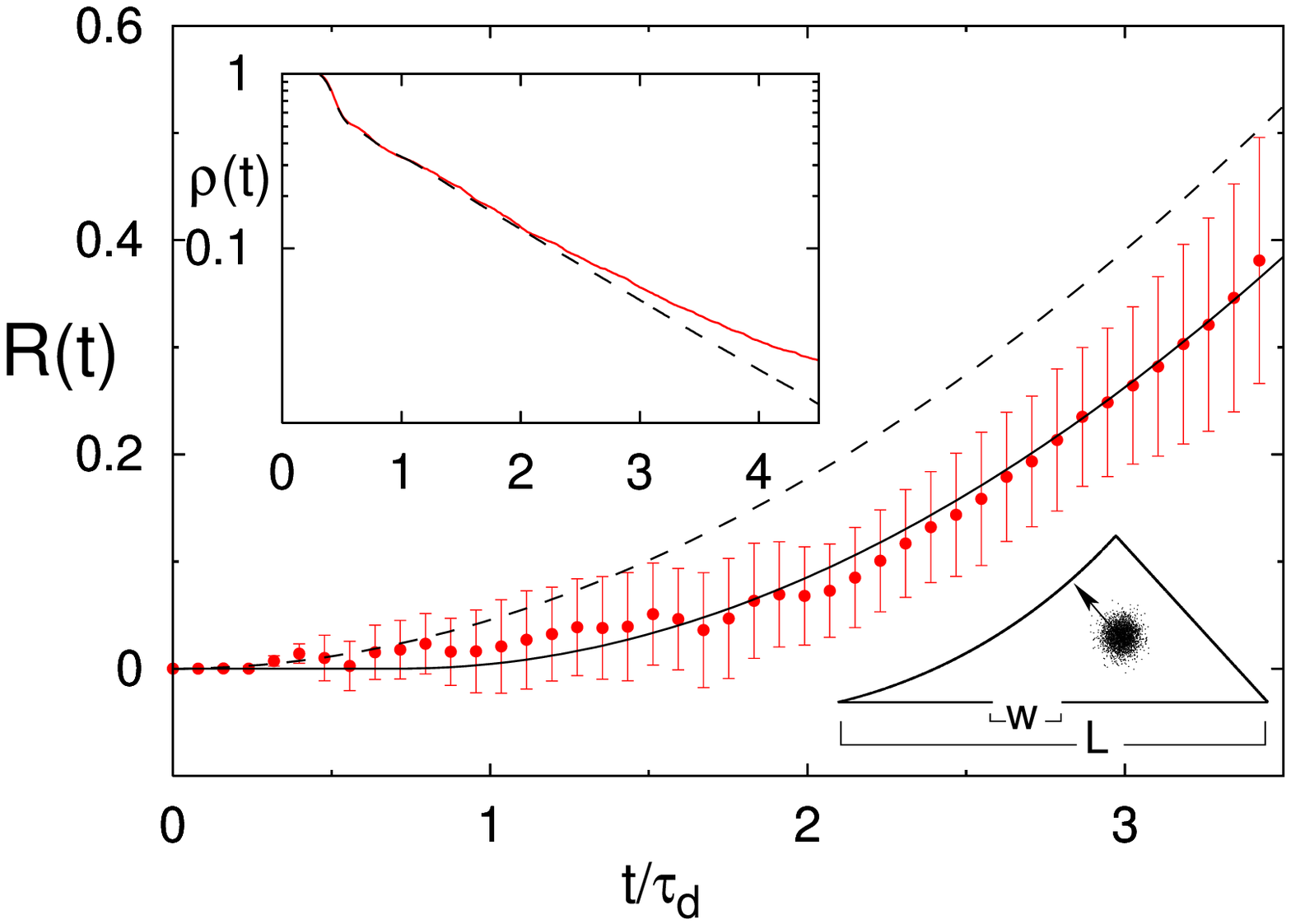}
\caption{(Color online) Averaged ratio  $R(t)$ between numerical quantum mechanical decay $\rho^{\rm
sim}_{\rm qm}(t)$ and classical decay
$\rho^{\rm sim}_{\rm cl}(t)$ 
(red symbols, the bars correspond to the variance after averaging, see text)
compared with corresponding semiclassical predictions based on the quadratic term in Eq.\,(\ref{eq:GOE}) 
(dashed line) and Eq.\,(\ref{eq:Ehrenfest}) (full line). 
Upper inset: $\rho^{\rm sim}_{\rm qm}(t)$ (red full line) and $\rho^{\rm sim}_{\rm cl}(t)$ (dashed)
for the wave packet shown in the
lower inset. Lower inset: Desymmetrized diamond billiard, defined as the fundamental domain of the area confined 
by four intersecting disks of radius $R$ centered at the vertices of a square of length $2L$ 
($L=100$, $R=131$) with opening $w=16$. The initial Gaussian wave packet shown is of
size $\sigma\!=\! 3$ and $\lambda_{\rm B}\!=\!3$.
The arrow marks the momentum direction. 
}
\label{fig:numi}
\efg

{\it Numerical simulation.--}
The leading-order quantum corrections in Eq.\,(\ref{eq:GOE}) and
(\ref{eq:GUE}) were confirmed by numerical simulations for graphs
\cite{ref:Puhlmann05}. Here we compare our semiclassical predictions
with quantum calculations of $\rho(t)$ based on the numerical
propagation of Gaussian wave packets inside a billiard, a setup much
closer to experiment. We chose the desymmetrized diamond billiard
(inset Fig.\,\ref{fig:numi}) \cite{ref:Goussev07} that is classically
chaotic ($\lambda^{-1}\simeq 3\tau_f$, with $\tau_f$ the mean free
flight time). Its opening $w$ corresponds to $N\!=\!10$ open channels and to $\tau_d\! \simeq\! 15 \tau_f$.  For the simulations, we reach $t_H \!=\!
10.6\tau_d$ implying $t^\ast\!=\! 3.3\tau_d$, $\tau_E^{\rm o}\!
\simeq\! 0.17 \tau_d$ and $\tau_E^{\rm c}\! \simeq\! 0.55 \tau_d$
(with $\mathcal L\!=\!\sqrt{A}$).

In the upper inset of Fig.\,\ref{fig:numi} we compare the decay, 
$\rho^{\rm sim}_{\rm qm}(t)$
(red full line), for a representative wave packet simulation with the corresponding classical, 
$\rho^{\rm sim}_{\rm cl}(t)$ (dashed line), 
obtained from an ensemble of trajectories 
with the same phase space distribution as the Wigner function of the initial quantum 
state. 
$\rho^{\rm sim}_{\rm cl}(t)$ merges into the exponential decay 
$ \exp(-t/\tau_d)$, and
$\rho^{\rm sim}_{\rm qm}(t)$
coincides with  $\rho^{\rm sim}_{\rm cl}(t)$ up to scales of $t^\ast$.
For a detailed analysis of the quantum deviations we consider the ratio 
$R(t)\!\equiv\! [\rho^{\rm sim}_{\rm qm}(t)\!-\!\rho^{\rm sim}_{\rm cl}
(t)]/\rho^{\rm sim}_{\rm cl}(t)$.
The red symbols in Fig.\,\ref{fig:numi} represent an average of $R(t)$ over 27 
different opening positions and initial momentum directions.
The dashed and full curve depict the semiclassical results based on 
the quadratic term in Eq.\,(\ref{eq:GOE}) (dominant for the $t/\tau_d$-range displayed) and
on Eq.\,(\ref{eq:Ehrenfest}). 
The overall agreement of the numerical data with the full curve indicates $\tau_E$-signatures. We note, however, that we cannot rule out other non-universal effects 
(e.g.\ due to scars  \cite{ref:Kaplan}, short orbits,
diffraction or fluctuations of the effective $\tau_d$ \cite{ref:Casati})
that may also yield time shifts. 
Furthermore the individual numerical traces $R(t)$ exhibit strong fluctuations (reflected in
 a large variance  in Fig.\,\ref{fig:numi}).
A numerical confirmation of the $\log(1/\hbar)$-dependence of $\tau_E$
 seems to date impossible for billiards.

 {\it Photo-ionization and -dissociation cross sections.--} 
Related to the decay problem are photo absorption processes where a
molecule \cite{ref:Schinke93} (or correspondingly an atom) is excited into a 
classically chaotic, subsequently decaying resonant state. In dipole approximation, 
the photo-dissociation cross section of the molecule excited from the ground state
$|g\rangle $ 
is $\sigma(\epsilon)\!=\! {\rm Im} {\rm Tr}\{ \hat{A} G(\epsilon) \}$,
where $G(\epsilon)$ is the retarded molecule Green function,
$\hat{A}\!=\! [\epsilon /(c\hbar \epsilon_0)] |\phi\rangle \langle \phi |$
and $|\phi\rangle\!=\!D|g\rangle$, with $D\!=\!{\bf d}\! \cdot \! \hat{{\bf e}} $ 
the projection of the dipole moment 
on the light polarization $\hat{\bf e}$.
The two-point correlation function of $\sigma(\epsilon)$ is defined as
\beq\label{eq:2-point-corr}
C(\omega)= \langle \sigma(\epsilon+\omega \Delta/2)
\sigma(\epsilon-\omega \Delta/2)\rangle_\epsilon / \langle{\sigma}(\epsilon)\rangle_\epsilon^2 - 1 \, ,
\eeq
where $\langle{\sigma}(\epsilon)\rangle_\epsilon 
\approx \pi(2\pi\hbar)^{-2}\! \int {\mathbf dr}{\mathbf dp} 
A_W({\mathbf r},{\mathbf p})\delta(\epsilon\!-\!H({\mathbf r},{\mathbf p} ))$ semiclassically,
with Wigner transform $A_W$ of $\hat A$. 
Previous semiclassical treatments of $C(\omega)$ \cite{ref:Agam00,ref:Eck00}
were limited to the diagonal approximation. To compute off-diagonal (loop) terms
we consider 
$ Z(\tau) \!=\! \int_{-\infty}^{\infty} d\omega e^{2\pi i\omega \tau}C(\omega)$
with $\tau\!=\! t/t_H$.
Semiclassically, $Z_{\rm sc}$ is again given by a double sum 
over orbits with different initial and final points. 
Due to rapidly oscillating phases from the action differences, 
only two possible configurations of those points contribute
 \cite{ref:Agam00}:
(i) orbits in a sum similar to 
Eq.\,(\ref{eq:rho-sc}) leading to a contribution as for $\rho_{\rm sc}(t)$;
(ii) trajectories in the vicinity of
a periodic orbit. Expanding around it, as in \cite{ref:Eck00},
leads to the spectral form factor $ K^{\rm open}_{\rm sc}(\tau)$
of an open system. From (i) and (ii) we have
$Z_{\rm sc}(\tau)\!=\!  K^{\rm open}_{\rm sc}(\tau) \!+\!  2\rho_{\rm sc}(\tau)$
for the time reversal case.
Up to second order in $\tau\!>\!0$ we find
$ K^{\rm open}_{\rm sc}(\tau)\!=\!{\rm e}^{- N\tau}(2\tau-2\tau^2)$ and
$\rho_{\rm sc}(\tau)\!=\! {\rm e}^{- N\tau}(1+N\tau^2/2)$ (Eq.\,(\ref{eq:GOE})),
Thereby
$Z_{\rm sc}(\tau)\!=\! e^{-N\tau} [2\!+\!2\tau\!+\!(N\!-\!2)\tau^2]$,
confirming a conjecture of \cite{ref:Gorin05}.
Its inverse Fourier transform yields the 
two-point correlation 
(with $\Gamma=2\pi\omega\tau_d/t_H$)
\beq
C_{\rm sc}(\Gamma)\!=\!\frac{4}{N}\!\frac{1}{1\!+\!\Gamma^2}\!\left[ 1 \!+\!
  \frac{1}{N}\frac{1\!-\!\Gamma^2}{1\!+\!\Gamma^2}
    \!+\!\frac{N\!-\!2}{N^2}\frac{1\!-\!3\Gamma^2}{(1\!+\!\Gamma^2)^2}\right]\! . 
\eeq
The first two diagonal terms agree with \cite{ref:Agam00};
the third term represents the leading quantum correction.

To conclude, we presented a general semiclassical approach to the
problems of quantum decay and photo cross-section statistics in open
chaotic quantum systems.

We thank I.~Adagideli, J.~Kuipers and C.~Petitjean for useful discussions and 
for a critical reading of the manuscript.
We acknowledge funding by DFG under GRK 638 and the A.~von Humboldt Foundation (AG).


\begin{thebibliography}{00}
\bibitem{ergodic}
We consider times $ \lambda t \gg 1 $ for a chaotic system 
with Lyapunov exponent $\lambda$ and a small opening such that 
$\lambda\tau_d\gg 1$, while 
the number $N$ of channels is still large.
\end{thebibliography}
\end{document}